
\input phyzzx
\PHYSREV
\hbox to 6.5truein{\hfill UPR-0525T }
\hbox to 6.5truein{\hfill September, 1992}
\title{PARTICLE WIDTH EFFECT ON $CP$-VIOLATING ASYMMETRIES}
\author{Jiang Liu}
\address{Department of Physics, University of Pennsylvania, Philadelphia,
         PA 19104}
$$ $$
\abstract
$CP$-violating asymmetries calculated from the Breit-Wigner
approximation for unstable particle propagators violate $CPT$.
A formalism satisfying $CPT$ invariance and unitarity is presented.
Applications are given to  $t$ decays.  For the decay
$t\to b\nu_{\tau}\bar{\tau}$ in a class of $CP$-violating
Higgs models the $CP$-violating asymmetry
 turns out to be $10^{-4}$ times smaller
than the previously reported results.

PACS\#: 11.30.Er, 13.20.Jf, 13.30.Ce.
\endpage

Final-state interactions play an important role in the determination
of $CP$ and $T$ violation.  A test for $CP$ violation is to compare
the partial decay rates of a particle and its antiparticle.  In this
case final-state interactions are necessary since in their absence
the partial decay rates are equal from $CPT$ invariance even if $CP$
is violated\rlap.\Ref\wolf{For a review see: L. Wolfenstein,
Ann. Rev. Nucl. Part. Sci. 36, 137 (1986).}

Examples of interest are those involving  an intermediate
unstable particle, in which the final-state interaction enters into the
standard calculation as the width of the unstable particle.  In this
case the Breit-Wigner approximation is usually introduced, under which
the $W$ gauge boson propagator is given in the unitary
gauge by
$$iD_{\mu\nu}(k)_{BW}={-i\over k^2-M_W^2+i\Gamma_WM_W}\Bigl[g_{\mu\nu}
-{k_{\mu}k_{\nu}\over M_W^2}\Bigr],\eqno(1)$$
where $M_W$ and $\Gamma_W$ are the $W$ mass and total width, respectively.
The subscript $BW$ refers to the Breit-Wigner approximation.

This approximation violates $CPT$ invariance and its effect can be
very large in calculating delicate quantities such as $CP$-violating
asymmetries in weak decays.

Consider the difference of the following partial rates of the $t$ decays
$$\Delta_{b\nu_{\tau}\bar{\tau}}\equiv \Gamma(\bar{t}\to\bar{b}
\bar{\nu}_{\tau}\tau)-\Gamma(t\to b\nu_{\tau}\bar{\tau})\eqno(2)$$
in the $CP$-violating Higgs model of Weinberg\rlap.\Ref\Winberg{
             S. Weinberg, Phys. Rev. Lett. 37, 657 (1976).}
We consider this example because (1) there is a considerable interest
 in these decays\rlap,\REFS\soni{G. Eilam, R. R. Mendel,
             R. Migeron, and A. Soni, Technion-Israel Inst. Tech. report,
             TECHNION-PH-92-10, (1992).}\REFSCON\gunion{
             R. Cruz, B Grz\c{a}dkowski, and J. F. Gunion,
             UC Davis report, UCD-92-15, (1992).}\refsend
and (2) in this case the $CPT$ violation effect due to the Breit-Wigner
approximation is most drastic.

The interaction between the charged-Higgs bosons and the fermions
of the model
is given by\Ref\albright{ C. H. Albright, J. Smith, and S.-H. Tye,
             Phys. Rev. D21, 711 (1980).}
$$L(H_{1,2}^+)={g\over\sqrt 2M_W}\sum_{i=1}^{2}H_i^+\Bigl[
     \bar{U}\Bigl(Y_iM_uKL+X_iKM_dR\Bigr)D-Z_i\bar{\nu}M_{\ell}R\ell
     \Bigr],\eqno(3)$$
where $M_u$, $M_d$ and $M_{\ell}$ are the mass matrices for the up
quarks, down quarks, and leptons, respectively. $K$ is the
Kobayashi-Maskawa matrix\rlap,\Ref\KM{M. Kobayashi and T.
             Maskawa, Prog. Theor. Phys. 49, 652 (1973).}
$L$ and $R$ are the helicity projection operators.
$X_i$, $Y_i$ and $Z_i$ are dimensionless parameters arising
from the diagonalization of the mass matrix of the charged scalars.
They depend on three mixing angles and one $CP$-violating
phase.

Neglecting the  contributions from the neutral Higgs sector\rlap,
\Ref\neutral{Interactions involving  a charged- and a neutral-Higgs
boson can provide a dominant contribution to the $CP$-violating
asymmetry provided
\splitout
$m_t\ge m_b+M_{H^+}$, where $H^+$ is
either $H_1^+$ or $H_2^+$ or both.  This possibility will not be discussed
in this paper.}
 the lowest order
$CP$-violating asymmetry $\Delta_{b\nu_{\tau}\bar{\tau}}$ is produced by
the interference of the tree-level
$W$ and  charged-Higgs exchange diagrams.  Assuming, for simplicity,
that the
contributions from  $H_2^+$ are negligible and that the final-state interaction
effect due to the charged-Higgs boson width is insignificant,
a straightforward  calculation shows that the Breit-Wigner approximation result
is
$$\Delta_{b\nu_{\tau}\bar{\tau}}={-g^4\over 512\pi^3}\Bigl({m_{\tau}
\over M_W}\Bigr)^2{\tilde{\Gamma}_W\over M_W}{m_t^3M_W^2Im(Y_1Z_1)\over
(M_W^2-M_{H_1^+}^2)^2+\Gamma_W^2M_W^2}F\Bigl({M_W^2\over m_t^2},
{M_{H_1^+}^2\over m_t^2}\Bigr),\eqno(4)$$
where
$$F(x,y)=\Bigl[{y\over x}-1\Bigr]\Bigl[(1-x)^2\ln{1-x\over x}
          -(1-y)^2\ln{1-y\over y}+(x-y)\Bigr],\eqno(5)$$
and
$$\tilde{\Gamma}_W=\Gamma_W-\Gamma(W\to \nu_{\tau}\bar{\tau})
   \approx {8\over 9}\Gamma_W.\eqno(6)$$
In reaching Eq. (4) we have ignored the $b$-quark mass and
assumed $\Gamma_WM_W\ll \vert M_W^2-M_{H_1^+}^2\vert$.
The dependence of $\Delta_{b\nu_{\tau}\bar{\tau}}$ on $\tilde{\Gamma}_W$
arises because the same final state rescattering does not contribute
to its $CP$-violating asymmetry\rlap.\Ref\Lincoln{L. Wolfenstein,
         Phys. Rev. D43, 151 (1991).}
An important feature of Eq. (4) is that $\Delta_{b\nu_{\tau}\bar{\tau}}$
is quadratic in $m_{\tau}$.  One power of $m_{\tau}$ comes from
$L(H_1^+)$ (see Eq. (3)).
The other is the price which one pays for
the interference of a vector-current (due to $W^+$) and a scalar-current
(due to $H_1^+$).

Under the Breit-Wigner approximation $\Delta_{b\nu_{\tau}\bar{\tau}}$
is the largest $CP$ asymmetry that can be produced by the standard
weak currents and the interaction given by Eq. (3).  The analogous
asymmetries of other $t$ decays are completely negligible:
the other semileptonic decays are  suppressed
by either $m_{\mu}^2$ or $m_e^2$, where $m_{\mu(e)}$ is the
muon(electron) mass;  the lowest order hadronic decays
conserve $CP$ if we ignore the small down-type quark masses.
It then follows that there would be no other $CP$-violating asymmetries
big enough to cancel $\Delta_{b\nu_{\tau}\bar{\tau}}$, and thus
the lifetime of $t$ would  not be the same as the
lifetime of $\bar{t}$, in violation of $CPT$ invariance.

$CPT$ violation occurred in the calculation discussed above is a
consequence of the Breit-Wigner approximation, which misses an important
final-state interaction phase.

Consider the $W$ propagator
in the unitary gauge.  The  tree-level
result is
$$iD_{\mu\nu}^{(0)}(k)={-i\over k^2-M_W^2}\Bigl[g_{\mu\nu}-
     {k_{\mu}k_{\nu}\over M_W^2}\Bigr].\eqno(7)$$
Parameterizing the one-loop $W$ self-energy as
$$\Pi_{\mu\nu}(k)=g_{\mu\nu}F_1(k^2)+(k^2g_{\mu\nu}-k_{\mu}
     k_{\nu})F_2(k^2),\eqno(8)$$
we find that the regularized (but unrenormalized) $W$ propagator
is
$$\eqalign{iD_{\mu\nu}^{(1)}(k)&
=-i\Bigl[(k^2g_{\mu\nu}-k_{\mu}k_{\nu})-M_W^2g_{\mu\nu}-\Pi_{\mu\nu}
         (k^2)\Bigr]^{-1}\cr
&={-i[1-F_2(k^2)]^{-1}\over
   k^2-[M_W^2+F_1(k^2)][1-F_2(k^2)]^{-1}}\Bigl[g_{\mu\nu}-
   k_{\mu}k_{\nu}{1-F_2(k^2)\over M_W^2+F_1(k^2)}\Bigr].\cr}\eqno(9)$$
In analogous to the optical theorem,
the unitarity of $S$-matrix implies a  relation between $Im\Pi_{\mu\nu}
\epsilon^{\mu}\epsilon^{\nu}\vert_{k^2=M_W^2}$ and $\Gamma_W$:
$$ImF_1(M_W^2)+M_W^2ImF_2(M_W^2)=-\Gamma_WM_W,\eqno(10)$$
where $\epsilon_{\mu}$ is the
$W$ polarization vector.
It then follows from the standard on-shell renormalization procedure
that
$$iD_{\mu\nu}^{(1)}(k)=
   {-iQ(k^2)\over k^2-M_W^2+i\Gamma_WM_W-\Sigma_R(k^2)}
   \Bigl[g_{\mu\nu}-{k_{\mu}k_{\nu}\over M_W^2-i\Gamma_WM_W
   +\Sigma_R(k^2)}\Bigr],\eqno(11)$$
where the mass renormalization correction is
$\Delta M_W^2=ReF_1(M_W^2)+M_W^2ReF_2(M_W^2)$ and
$$\eqalign{&Q(k^2)=[1-F_2(k^2)+F_2(M_W^2)]^{-1},\cr
&\Sigma_R(k^2)=[F_1(k^2)-F_1(M_W^2)]+M_W^2[F_2(k^2)-F_2(M_W^2)]\cr}
  \eqno(12)$$
are the renormalized weak charge correction and self-energy,
respectively.
Complications in the longitudinal part of the propagator
are usually neglected in the literature, for in most of the practical
cases the
$k_{\mu}k_{\nu}$ term  either vanishes (if it couples to a conserved
current) or is very small
(if it couples to fermions with small masses).

Compared to the Breit-Wigner approximation (Eq. (1)), the result given
by Eq. (11) has an important difference:  the width term
$i\Gamma_WM_W$ is always in company with the mass term $M_W^2$
everywhere in the formula,
not just in the denominator of the propagator.  Other modifications
include (1) an over all screening factor $Q(k^2)$
for the weak charge, and (2) a renormalized self-energy $\Sigma_R(k^2)$.
Both $Q(k^2)$ and $\Sigma_R(k^2)$
 can have a final-state interaction phase, but
their effects are likely to be less significant
as they  vanish at the pole  (see below).
The position of the pole of the propagator
remains the same, i.e.,
$$ k^2=M_W^2-i\Gamma_WM_W,\eqno(13)$$
and the relations $\Sigma_R(k^2)\vert_{k^2=M_W^2-i\Gamma_WM_W}=0$
and $Q(k^2)\vert_{k^2=M_W^2-i\Gamma_WM_W}=1$
hold to the order of interest to us in
the perturbation expansion\rlap.\Ref
\width{The effect of particle width on the definition of particle mass becomes
slightly more complicated in higher orders of perturbation expansion:
\splitout
A. Borrelli, M. Consoli, L. Maiani and R. Sisto, Nucl. Phys. B333, 357 (1990);
A. Borrelli, L. Maiani and R. Sisto, Phys. Lett. B244, 117 (1990);
A. Sirlin, Phys. Rev. Lett. 67, 2127 (1991); R. Stuart, Phys. Lett. B262,
113 (1991);  G. Valencia and S. Willenbrook, Phys. Rev. D40, 853 (1990);
Phys. Lett. B259, 373 (1991).}

Turn to the calculation of the $CP$-violating asymmetries in the
$t$ decays.  For the case at hand, effects arising from the weak charge
screening factor $Q(k^2)$ and the self-energy
 $\Sigma_R(k^2)$ are negligible, and hence
to a good approximation the $W$ propagator can be written as
$$iD_{\mu\nu}(k)\approx {-i\over k^2-M_W^2+i\Gamma_WM_W}\Bigl[
       g_{\mu\nu}-{k_{\mu}k_{\nu}\over M_W^2-i\Gamma_WM_W}\Bigr].
    \eqno(14)$$
The phase in the longitudinal part of the propagator is crucial
in restoring $CPT$ invariance of the calculation.  It introduces
a correction to $\Delta_{b\nu_{\tau}\bar{\tau}}$ given by
Eq. (4).  The correction has the same amplitude as Eq.(4)
but with the opposite sign.
As a consequence, the $CPT$-violating result (Eq. (4)) cancels
completely, as it should.

It is not totally
surprising that the final-state interaction phase in the
$k_{\mu}k_{\nu}$ term of the $W$ propagator has such a big effect.
It has been noticed by the previous studies (Refs. \soni,\ \gunion\ )
that under the Breit-Wigner approximation the interference term of the
squared amplitudes vanishes for an on-shell $W$. This does not happen to
the correction terms however, even though they appear to be suppressed by a
factor $\Gamma_W/M_W$.

Because of the large cancellation imposed by $CPT$ invariance,
the size of the $CP$-violating asymmetries induced by the mechanism
discussed above turns out to be purely academic.
The order-of-magnitude of the leading terms is approximately
$$\eqalign{
\Delta_{b\nu_{\tau}\bar{\tau}}&=-\Delta_{bc\bar{s}}\cr
&\sim {-g^4\over 512\pi^3}\Bigl({m_{\tau}\over M_W}\Bigr)^2
\Bigl({m_c\over M_W}\Bigr)^2{\tilde{\Gamma}_W\over M_W}
{m_t^3M_W^2\vert Y_1\vert^2Im(Y_1Z_1)
\over (M_W^2-M_{H_1^+}^2)^2+\Gamma_W^2M_W^2}
f\Bigl({M_W^2\over m_t^2},
{M_{H_1^+}^2\over m_t^2}, {m_{\tau}^2\over m_t^2},
{m_c^2\over m_t^2}\Bigr),\cr}\eqno(15)$$
where $\Delta_{bc\bar{s}}=\Gamma(\bar{t}\to\bar{b}\bar{c}s)
-\Gamma(t\to bc\bar{s})$,
$m_c$ is the $c$-quark mass, and $f$ is a function of order unity
or less and symmetric under the exchange of $m_{\tau}$ and $m_c$.
Although the calculations of the decay rates $\Gamma(t\to
b\nu_{\tau}\bar{\tau})
$ and $\Gamma(t\to bc\bar{s})$ are very different, there is a complete
correspondence imposed by $CPT$ invariance when we calculate the
rate difference.
Compared to Eq. (4), the unitarized and $CPT$ invariant
result (Eq. (15)) has an
additional suppression factor $\sim \vert Y_1\vert^2m_c^2/M_W^2$ and
is about $10^{-4}$ times smaller.

In general,
neglecting the phase in the $k_{\mu}k_{\nu}$
part of the $W$ propagator will have
an important implication for $CP$ asymmetries
generated from interactions
involving  an interference  that requires a small mass insertion.
The effect is expected
to be small
for interactions\Ref\Hewett{G. Eilam, J. L. Hewett, and A.
         Soni, Phys. Rev. Lett. 67, 1979 (1991); {\it{ibid}} 68, 2103
         (1992); J. M. Soares, Phys. Rev. Lett. 68, 2102 (1992).}
involving currents of the same type,
unless the final-state interaction phase factors.

The propagator of an unstable fermion with mass $m$ and
width $\Gamma$ is
$$iS(k)={i\over k\!\!\!/-m+i\Gamma/2-\Sigma_R(k^2)}\approx
       i{k\!\!\!/+m-i\Gamma/2\over k^2-m^2+i\Gamma m},\eqno(16)$$
where $\Sigma_R(k^2)$ is the renormalized self-energy of the fermion.
The approximation in the second step of Eq. (16)
corresponds to neglecting $\Sigma_R(k^2)$.
The phase factor $i\Gamma/2$ in the numerator has so far been ignored
in the literature\rlap.
      \Ref\fermi{
      M. Nowakowski and A. Pilaftsis,
      Mod. Phys. Lett. A4, 821 (1989); Z. Phys. C42, 449 (1989);
      Mod. Phys. Lett. 6A, 1933 (1991);  A. Pilaftsis, Z. Phys. C47,
      95 (1990).}
Its consequence should be examined carefully.  Little change is expected
from an unstable scalar, whose propagator is  the same
as the standard Breit-Wigner form if one ignores its
renormalized self-energy.

In conclusion, we have shown that the standard Breit-Wigner
approximation for an unstable particle violates $CPT$.
The effect can be large in calculating delicate quantities
such as $CP$-violating asymmetries in some theoretical models.
Formalisms satisfying $CPT$ invariance and unitarity
are derived.  For the $CP$-asymmetry discussed
in more detail in this paper,  their applications
show that the result is about $10^{-4}$
times smaller than that obtained from the Breit-Wigner approximation.

\ack

I wish to thank M. Cveti\v c, B. Kayser, P. Langacker and L. Wolfenstein
for valuable discussions and suggestions.  I would also like to
thank Aspen Center for Physics for hospitality during the initial
stage of this work.  This research was supported in part by the
U. S. Department of Energy under the contract DE-ACO2-76-ERO-3071
and the SSC National Fellowship.
\endpage
\refout
\end